# Enhanced reflectivity change and phase shift of polarized light: double parameter multilayer sensor

*Ilze Aulika\*, Martins Zubkins, Jelena Butikova, Juris Purans*
Institute of Solid State Physics, University of Latvia (ISSP UL), LV-1063, Riga, Latvia, email: *ilze.aulika@cfi.lu.lv*

Here the concept of point of darkness based on polarized light phase difference and absorption of light is demonstrated by simulations using low refractive index and extinction coefficient semiconductor and dielectric, and high refractive index non-oxidizing metal multilayer thin film structures. Several multilayer sensor configurations show great sensitivity to thickness and refractive index variation of the detectable material by measuring the reflectivity ratio $\Psi$ and phase shift $\Delta$. Focus is on such multilayers, which have sensitivity to both parameters ($\Psi$, $\Delta$) in the visible spectral range, thus, opening the possibility for further research on a new biomedical sensor development with enhanced double parameter sensing.

## 1. Introduction

Biomedical sensor development for healthcare and diagnostics is a very rapidly growing market, which was valued at USD 22.4 billion in 2020 and is expected to expand at a compound annual growth rate (CAGR) of 7.9% from 2021 to 2028[1]. In general, there are two types of biosensors - sensor patch and embedded devices, which use different detection mechanisms, e.g., electrochemical, optical, piezoelectric, thermal, and nanomechanical biosensors. Among all these technologies, the optical sensors have several advantages in the number of different receptor types and, moreover, permit the integration of different sensing mechanisms. Recent and very elaborated review (covering nearly 250 research articles) about optical biomedical sensors based on surface plasmon resonance (SPR), evanescent wave fluorescence, reflectrometric (sometimes called also refractive), bioluminescence and several others, showed the huge potential for the broad use of these sensors in point-of-care testing devices and early-stage disease prediction[2]. Reflectometric biosensors[3] are very attractive because they are relatively simple with respect to SPR and other methods, they are very sensitive, time-resolved and label-free techniques. Under reflectometric principle of biosensing we can find such methods as reflectometric interference spectroscopy[4], interferometric reflectance imaging[5], and spectroscopic ellipsometry[6] (SE), where the main sensing response is to a slight change in layer thickness *d* seen as a shift of the interference pattern within the wavelength domain. The analyte concentration is determined by variations in the amplitude and phase of polarized light due to changes in the refractive index *n* and thickness of an adsorbed layer of the analyte[1, 4-6] (**Figure 1**).

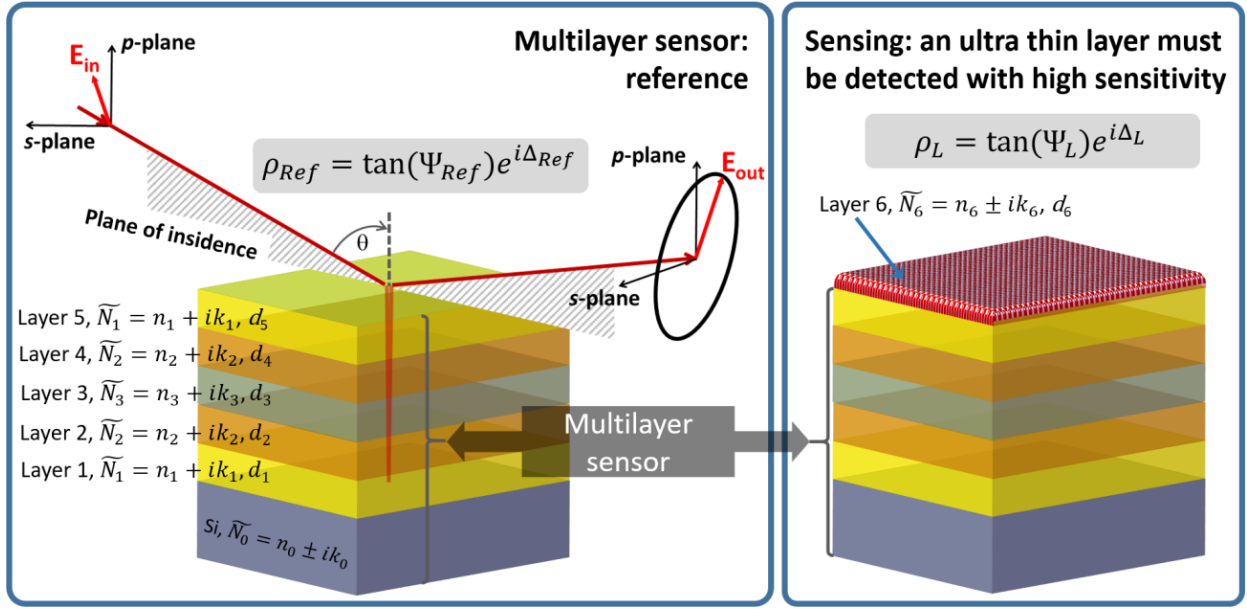

**Figure 1.** Schematic illustration of the sensory device based on reflectivity measurement set-up with the capability to measure the ratio of the amplitude and the phase shift.

### 1.1 Spectroscopic ellipsometry

SE is a very powerful method for sensing very thin layers. Many of the commercially available SE instruments measure the main ellipsometric angles ($\Psi$, $\Delta$) with 0.001° resolution, thus SE can measure the sub-angstrom thickness change[7] and can detect the presence of very small amount of substances. For example, experimentally, it was reported detection limits of immunoglobulin G (IgG) 15 ng/ml, 0.1 ng/ml of Hepatitis B surface antigen, 0.01 ng/ml α-fetoprotein and 10 amol/ml in DNA-hybridization studies[6-10]. SE has been demonstrated to be a very promising sensor[6-10]. There are first signs towards miniaturization of SE and SPR[11-13], and active research on metasurfaces for applications as active optical components[14] in miniature optical sensor and devices. It is expected that the optical biosensor development and market share with respect to other biosensor technologies is going to increase very rapidly in the nearest decades[1,2,15].

In ellipsometry, the two main ellipsometric angles (Ψ, Δ), functions of wavelength $\lambda$ and incident angle of light θ, are defined by

$$\rho \equiv \tan(\Psi)e^{i\Delta} = \frac{|\tilde{r}_p|}{|\tilde{r}_s|} = \frac{E_p^{out}/E_p^{in}}{E_s^{out}/E_s^{in}} \quad (1),$$

where $\tilde{r}_p$ and $\tilde{r}_s$ are Fresnel's coefficients for *p* (parallel) and *s* (perpendicular) polarized light, $E_p^{out}, E_p^{in}, E_s^{out},$ and $E_s^{in}$ represent *p* and *s* components of the incident and reflected electric fields of the light beam[16]. TanΨ is the amplitude ration upon reflection and $\Delta = \delta_p - \delta_s$ is the phase shift induced by the reflection between the *p*-wave and the *s*-wave. The ratio $\rho$ is related to the optical properties of the material or materials under investigation, namely, the complex refractive index $\tilde{N} = n + ik$, where *n* - is the index of refraction and *k* - extinction coefficient. The high sensitivity of this technique is derived from the fact that the measured (Ψ, Δ) is a relative measurement of the change in polarization (a ratio or difference of two values rather than the absolute value of either). It explains robustness, the high accuracy, and the reproducibility of the technique.

### 1.2 Other aspects of reflectance methods

Exploiting further the reflectance methods for biomedical sensor applications like SPR, reflectance and transmittance, it is important to develop such sensors which have high surface sensitivity: when any material of very low thickness (~0.1 nm) or of very small amount with different (*n*, *k*) values (with respect to the sensor) is present on the surface of the sensor, is detected by measuring reflectance, transmittance, phase shift, wavelength or incident angle shift. In the recent years much attention has been paid to the point of darkness[17,18,20] based on perfect absorption of light and phase singularity. The absence of reflection can be achieved at the Brewster angle $\theta_B$. Such suspension of the reflected light can be achieved in particular cases: ideal optical systems, prism-coupled SPR, the light reflection from a single interface[6], coherent absorption and parity-time metamaterial. Different subwavelength nanostructures also have been proposed for the realization of topological darkness[17,19]. Recently, it was reported on dispersion topological darkness at multiple wavelengths and polarization states using a three-layered structure where the top layer was nano-patterned[19]. Also, simpler systems were reported, where the point of darkness is achieved by the multistack made of, e.g., Ag (20 nm)/methyl methacrylate MMA (522 nm)/Ge (10 nm)/Ag (80 nm)[20]. For our knowledge, there are relatively many publications for multilayer structures with the point of darkness at NIR and IR wavelength range[19-25], and just few related to VIS spectra range[25,26]. It is since the high efficiency absorption at specific $\lambda$ is mainly due to the Fabry-Perot resonances in the non-absorbing dielectric middle layer, which result in trapping of the resonant light in the middle layer and thus enhanced absorption efficiency. The light trapping (multiple internal light reflecting) is provided by materials with high ($n$, $k$) values below and on the top of dielectric layer. This system can be imagined also as a kind of light guide where the internal layer has $n$ values lower with respect to the outer layer's $n$ values.

In case of prototyping of systems with point of darkness, the complete suppression of reflection is practically not possible due to sensor fabrication errors (disorder, inhomogeneity, thickness variation etc.), complex nano-patterning processes for metamaterial fabrication, natural oxidation of some metal layers, and polymer layer degradation under humidity and UV light. Interestingly, in case of multilayer stacks there are many studies devoted for the zero reflection (or perfect absorption), but not so many are dealing with the phase singularities and shifts. Probably it is related to the fact that it is easy to develop and do numerical simulation for the perfect absorption of light using, e.g., Fabry-Perot resonances or SPR approach, but both zero reflection and the phase sensitive sensors is a challenge. Phase-sensitive optical biosensing is of great importance in development of single molecule and low concentration solution studies[2,6,17,18,20]. The SPR approach, where the key parameters are the shift in the wavelength and/or incident angle, is extensively studied[14, 17, 18, 28-34], while change in reflectivity and phase shift is relatively less studied[6,17,18,20]. Some phase sensitive sensor systems are summarized in the **Table 1**, comparing technologically more challenging sensors with sensors fabricated with the simpler processes.

One possibility to achieve both (reflectivity and phase sensitivity) is by introducing into the multilayer stack a highly absorbing ultrathin layer, e.g., Ge[19,20, 27]. In this work, we propose an alternative: very thin multilayer system made of 5 films containing semiconductor and dielectric materials with low ($n$, $k$). By varying the thickness of the films and choosing semiconductors with lower refractive index with respect to Ge material, and combining them with dielectrics and metals in a multilayer stack, it is possible to obtain relatively broad spectral range for point of darkness, thus "adjusting" the sensor design to the easier, custom, more convenient or cheaper prototyping. This complementary approach offers a possibility to develop new and sustainable biomedical sensors, which can be deposited by common vacuum thin film technologies used in semiconductor industries with no need of lithography and etching processes. Moreover, this would be much simpler sensor configuration, where only single wavelength and single incident angle could be used.

**Table 1.** Some phase sensitive sensor description, fabrication technology, spectral range or λ and θ of work, detected material during sensor test and measured parameters (difference of the main SE angles dΨ and dΔ).

| Sensor description | Fabrication technology | Spectral range, nm | Detected material | Parameter | Ref. |
|---|---|---|---|---|---|
| **Plasmonic metamaterial; topologically protected darkness** | | | | | |
| SE of Au array with the graphene: average size of the dots 110 nm, separation 140 nm | High complexity: • e-beam lithography • e-beam evaporation • wet-etching • micromini • surface functionalization | Point of darkness at: • λ = 603 nm • θ ≈ 69° | Graphene hydrogenation; mass sensitivity < 10 fgmm$^{-2}$ | dΨ = 34° dΔ = 44° | [17] |
| SE of Au array with the graphene: average size of the dots 135 nm, separation 140 nm | | Point of darkness at: • λ = 710 nm • θ ≈ 53° | 10 pM streptavidin solutions in 10 mM phosphate-buffered; sensitivity 1-4 molecules per nanodot | dΨ = 25° dΔ = 25° | |
| Spectrometer and fringe pattern (for relative phase measurement) capture with color CCD of Au nanodiscs with 86 nm diameter and 20 nm hight on Cr/Poly(methyl methacrylate)/ borosilicate glass slide | Medium complexity: • thermal evaporation of Au, Cr • etching • tape-stripping • surface functionalization | Point of darkness at: • λ = (640-650) nm • θ ≈ 55° | Ethylene glycol 1) 2.5% ($n$=1.355) 2) 5% ($n$=1.360) 3) 60 kD protein NeutrAvidin (NA) | Relative phase shift from 1) ~170° and to 2) ~145°, ~25° phase shift for 0.005 refractive index difference, 3) Sensitivity 200–275 NA molecules per nanodisc | [18] |
| **Single surface and multilayer systems** | | | | | |
| θ = 45° dual-drive symmetric photoelastic modulator ellipsometer with microfluidic system based on Si/PDMS/prism | Medium complexity: • SiO$_2$ removal from Si • hydroxylation and amination • microfluidic flow fabrication (lithography) • surface functionalization | Biolayer θ$_B$ = 69°, when the θ = 55.07° for prism at λ = 650 nm | Human immunoglobulin IgG in the concentration range of 15-1000 ng/ml; limit of detection is 15 ng/mL | No sensitivity to Ψ. Simulation data: • Δ=-2.896$d$+177.583 for $n$ = 1.330 • Δ=-2.390$d$+178.013 for $n$ = 1.334 • Δ=-2.023$d$+178.300 for $n$ = 1.338 Additional data | [6] |

| | | | | added from the authors of this work:<br>• $d\Delta \approx 3°$ for 1 nm thickness change<br>• $d\Delta \approx -0.132 - 2.711d$<br>• $d\Delta \approx 66 - 50n$ | |
|---|---|---|---|---|---|
| SE of Ag/MMA/Ge/Ag on Si | Low complexity:<br>• thermal evaporation of Ag, Ge<br>• spin-coating of MMA<br>• surface functionalization | Point of darkness at:<br>• $\lambda = 708$ nm<br>• $\theta = 65°$ | 1 pM streptavidin with $\sim 11.8 \times 10^5$ $M^{-1}$ concentration | $d\Psi \approx 5.9°$<br>$d\Delta \approx 33°$ | [20] |

## 2. Modelling

The sensor and optical model for Si substrate with 5-6 layers is shown in the **Figure 1**. Main ellipsometric formula can be written with the following expressions:

- Reference samples defined as a clean sensor

$$\tan\Psi_{Ref}\exp(i\Delta_{Ref}) = \rho_{Ref}(\widetilde{N_0}, \widetilde{N_1}, \widetilde{N_2}, \widetilde{N_3}, d_1, d_2, d_3, d_4, d_5, \lambda, \theta)$$

- Sensor with some ultra-thin layer on the top

$$\tan\Psi_L\exp(i\Delta_L) = \rho_L(\widetilde{N_0}, \widetilde{N_1}, \widetilde{N_2}, \widetilde{N_3}, \widetilde{N_6}, d_1, d_2, d_3, d_4, d_5, d_6, \lambda, \theta)$$

were $d$ is the thickness of the film/layer. Each layer is characterized by its own $\widetilde{N}$ function and $d$, what is indicated with the numbering from 0 to 6. The **Figure 1** presents an example for the sensor made of 5 layers on Si, but this formalism can be applied to any other number of layers also on the substrates different from Si. The optical properties of the Layer 1 are equal with optical properties of the Layer 5, and also the Layer 2 and Layer 4 optical properties are identical, only thickness is different. Such "fiber" sandwich structure, where $\widetilde{N_0} > \widetilde{N_1} > \widetilde{N_2} > \widetilde{N_3}$, the electric field coupling and enhancement is facilitated [28,29]. Moreover, a great increase in phase shift can be obtained, thus providing a sensor with two sensitive parameters - reflectivity and phase.

The zero reflection and phase shift are the functions of wavelength $\lambda$, incident angle of light $\theta$, and polarization state of light. This concept can be written and analyzed using main ellipsometry formula (1):

(1) Phase shift Δ between phases $\delta_p$ and $\delta_s$ for *p* and *s* polarized light, which can reach values equal or near to -180° or +180° or alternatively equal or near to 0° or 360°; practically it is a situation, when the reflected waves are out of phase;

(2) Ratio of the complex Fresnel's coefficients $\tilde{r}_p$ and $\tilde{r}_s$ for *p* and *s* polarized light reaches zero $\tan\Psi = |\tilde{r}_p/\tilde{r}_s| = 0$, which is the case of perfect absorption of *p*-polarized light.

Typically, where Ψ reaches minimum or maximum values due to interference phenomena in thin films, the Δ might exhibit abrupt phase jumps. It should be noticed that it is not directly possible to use the abrupt Δ jump for sensing, thus the close to zero reflection condition and close to abrupt Δ jump measurements are sufficient for sensing applications.

The variation in Ψ and Δ is calculated as a difference in Ψ and Δ before and after the presence of another 0.1 nm thin layer on the top of the sensor (**Figure 1**): $d\Psi = \Psi_{Ref} - \Psi_L$ and $d\Delta = \Delta_{Ref} - \Delta_L$, where $\Psi_{Ref}$ and $\Delta_{Ref}$ are for the clean sensor as a reference, and $\Psi_L$ and $\Delta_L$ are in case the sensor has 0.1 nm thin layer on the top. The thin layer is simulated as polymethyl methacrylate (PMMA) material. The thickness of 0.1 nm can be considered thin enough to simulate sensor sensibility since dimension of organic molecules is in the range of (3 - 20) nm. For example, to reach the difference in Ψ and Δ for the sensor presented in the paper of K. V. Sreekanth *et al.*[20], for ideal multilayer structure the thickness of polymer layer on the top of the sensor must be at least (0.2 - 0.3) nm to obtain ~30° difference in Δ and at least (20 - 30) nm thick to reach ~6° difference in Ψ. The variation in the thickness are always accompanied with the large variation in Δ. In case of (20 - 30) nm polymer layer on the sensor, the variation in Δ should reach values up to 150°. Small variations in Δ reported in the work of K. V. Sreekanth *et al.*[20] probably are related to the biotin layer attached to the surface of the sensor to promote the streptavidin adhesion. The biotin and streptavidin probably have similar (*n*, *k*) values, thus reducing the variation in Δ with the presence of streptavidin on the biotin layer.

The multilayer sensor system presented in this paper is inspired by analysis and modelling of Ag/MMA/Ge/Ag system[20]. We found that for this system, for the given thicknesses the greatest variation in dΨ and dΔ is in UV region rather than in IR spectral range. This sensor, in principle, is an asymmetric Fabry-Perot configuration with large amount of optical loss Ge thin film thinner than the wavelength of light[20,27]. The high losses Ge supports the creation of "narrow" spectral zones with minimum values of Ψ and sharp Δ change, lowers total amplitude of Δ, and lowers the variation of dΨ and dΔ. This argument will be discussed in detail in a separate article. In case of phase sensitive sensor development working in visible spectral range, the low loss semiconductors could be considered to build an effective point of darkness.

On the other hand, since not always it is possible to develop optical sensors working with near UV or IR light, and not always it is convenient to work with oxidizing metals as Ag, other materials are proposed in this work: Ag is substituted with Au, and instead of high refractive index Ge semiconductor (n = 5.74 @ 600 nm), we propose low refractive index semiconductors, like YHO (n ≈ 2.28 @ 600 nm) or $ZnO_x$ (1.67 @ 600 nm)[23]. For the thicknesses of Au, YHO or $ZnO_x$, and $SiO_2$ smaller than the λ results in the 1) rise of dΨ and dΔ variation, 2) shift of the sensitive region from near UV to the visible range, 3) and in shift of sensitive incidence angles of light from ~70° to ~40°. The alternative multilayer proposed in this paper is made of 5 layers on Si (**Figure 1**): $Au(d_1)/YHO(d_2)/SiO_2(d_3)/YHO(d_4)/Au(d_5)$. Some discussion will be devoted also for the case, where even lower refractive index material with respect to YHO is used, such as $ZnO_x$. The impact of the thickness $d_1, d_2, d_3, d_4, d_5, d_6$ variation will also be discussed.

In the simulations we use the complex dielectric dispersion curves obtained from WOOLLAM RC2 SE investigations for YHO and $ZnO_x$[28] thin films developed at the Thin Film Laboratory of ISSP UL. The fabrication process and characteristics of $ZnO_x$ thin films can be found in our reference[35]. The YHO thin films were fabricated on soda-lime glass using the vacuum PVD coater G500M (Sidrabe Vacuum, Ltd.). YHO films were deposited by reactive DC magnetron sputtering from Y (purity 99.95 %) target in an Ar (99.9999%) and $H_2$ (99.999%) atmosphere ($H_2$ to Ar gas flow ratio of 2:3). The sputtering pressure was varied from 3 mTorr up to 20 mTorr. In this work, YHO sputtered at 3 mTorr obtained (*n*, *k*) dispersion curves are used (**Figure 2**) since such films are not completely oxidized and have *k* > 0 in the spectral range from UV to NIR; YHO sputtered at higher pressure exhibit photochromic properties and *k* > 0 only above optical band gap.

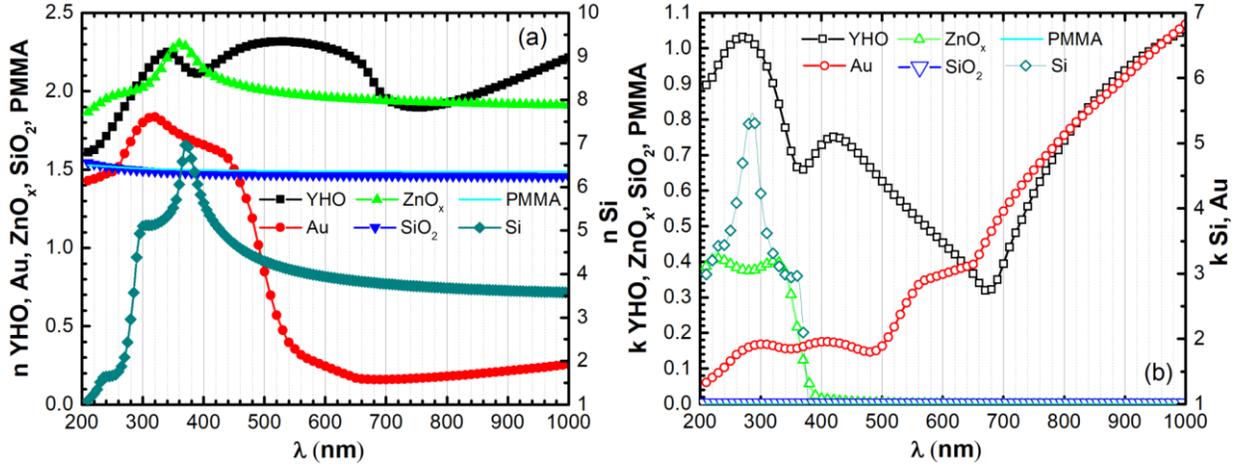

**Figure 2.** Dispersion curves of refractive index *n* (a) and extinction coefficient *k* (b) used in the modelling.

The dispersion curves for Si, $SiO_2$ and Au were selected from the WOOLLAM CompleteEASE software optical constant library. The (*n*, *k*) wavelength dependence of all materials used in simulations are present in the **Figure 2**. The simulations are performed using CompleteEASE by modelling the main ellipsometric angles ($\Psi$, $\Delta$) as a function of incident angle $\theta$ and photon energy $E = hc/\lambda$, where *h* - Planck's constant, and *c* - speed of light, and with COMSOL to obtain detailed maps of reflectivity and phase shifts in variation of the sensor thicknesses and incident angles of light to find the best combination of the thicknesses for the highest sensitivity. The modelling is performed in the spectral range of (1 - 5) eV or (248 - 1240) nm with two different steps: 0.02 eV or 1 nm, 0.0028 eV or 0.1 nm, and 0.05 nm. No modelling was performed for steps finer than 0.01 nm to be consistent with spectral resolution of common existing spectroscopic methods (e.g., WOOLLAM RC2 and similar SE wavelength resolution at UV-VIS is ~1.0 nm; Cary 7000 universal measurement spectrophotometer wavelength resolution at UV-VIS < 0.048 nm). The incident angle of light was varied in the range of (30.00 - 75.00)° with 2.0°, 0.5° and 0.1° steps.

### 3. Results

The modelling shows that the sensor has quasi zero reflection in the broad spectral range of (540 - 350) nm (or 2.3 - 3.5 eV, **Figure 3a**) for the following configuration: Si/Au (30 nm)/YHO (12 nm)/$SiO_2$ (35 nm)/YHO (15 nm)/Au (6 nm). This configuration gives the greatest variation in $\Delta$. More detailed spectra at $\theta = [30.0°; 50.0°]$ with the 0.5° step is illustrated in the **Figure 3b**. Sharp $\Delta$ shift of about 340° with quasi zero reflection is observed in the spectral range of ~ (505

- 300) nm. Also, at the shorter wavelengths of UV similar values can be seen (**Figure 3a**). In case of pure reflectivity measurements, the sensor is sensitive in the visible light spectral range of ~ (510 - 535) nm reaching d$\Psi$ values up to 0.22° and -0.12°, what is of 1) one order more for NIR and 2) approximately two times higher values for UV with respect to the sensors containing Ge[20]. In case of phase singularities, the largest d$\Delta$ of 225° is observed at $\lambda$ = 523.7 nm and at $\theta$ = 38.9° for the spectral scan (**Figure 4a**) and at d$\Delta$ = ±130° for $\theta$ = [38.7°; 39.0°] scan (**Figure 4b**). These d$\Delta$ values are 2-3 times higher with respect to sensors containing Ge[20]. A simple Si substrate biosensor[6] simulation shows that d$\Delta$ is only about 3° in case of 1 nm thin biolayer on the top of Si, and experiment evidenced no sensitivity to $\Psi$.

The biggest problem of the phase sensitive sensor development is the selection of the proper $\lambda$ and $\theta$ region due to the abrupt $\Delta$ jumps[19]. Even if very high d$\Delta$ is observed at particular $\lambda$ and $\theta$, it does not mean it will be possible to measure broad range of different thicknesses $d_6$ for the Layer 6 to be detected. With increase of $d_6$, the $\Delta$ and d$\Delta$ abrupt jump is shifting towards shorter $\lambda$ in case of $\lambda$ scan at fixed $\theta$, and towards smaller $\theta$ in case of $\theta$ scan at fixed $\lambda$ (**Figure 5**). For this reason, $\lambda$ and $\theta$ should be chosen near the point of darkness.

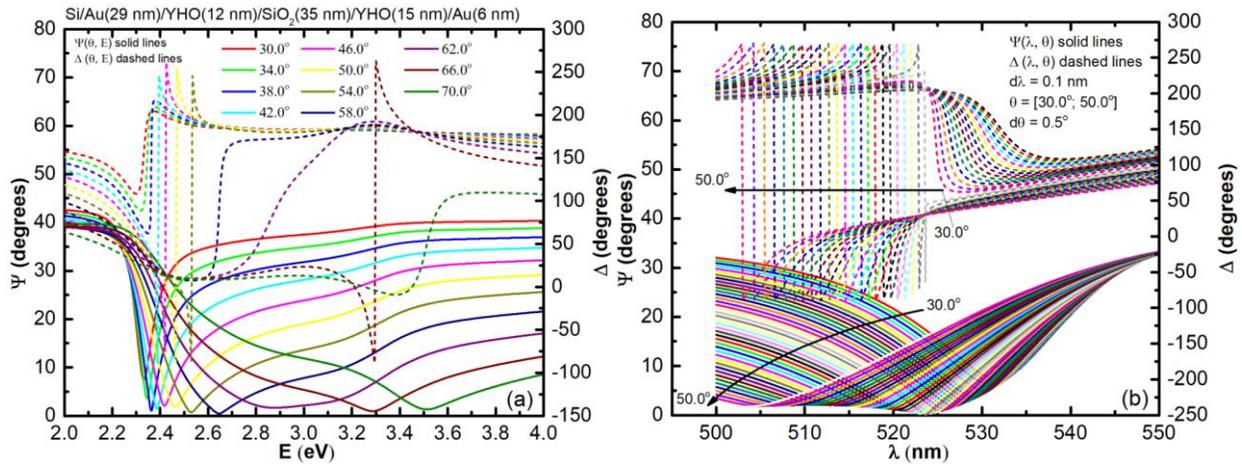

**Figure 3.** Main ellipsometric angles ($\Delta$, $\Psi$) as a function of photon energy (a) or wavelength (b) with (a) 0.02 eV and (b) 0.1 nm step. The incident angle is varied from 30° up to 70°.

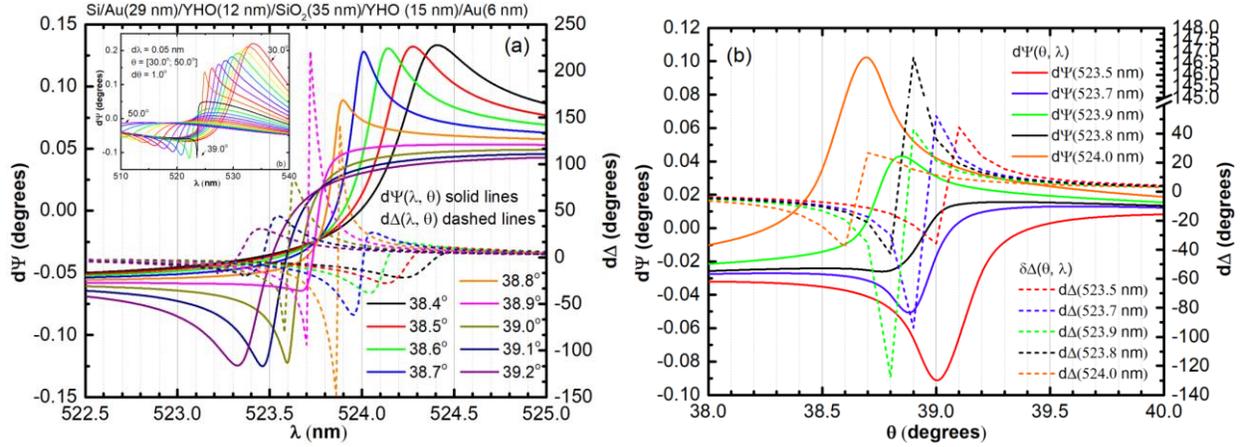

**Figure 4.** Two spectral sections (a, b) and three incident angle sections (c, d) Difference of Δ, Ψ as a function of the $\lambda$ (a) and $\theta$ (b) (dΔ, dΨ). The $\lambda$ is varied with the step of 0.05 nm, and $\theta$ with the step of 0.1°.

Depending on selected ($\theta$, $\lambda$), it is possible to obtain sensors sensitive 1) only to Ψ, 2) only to Δ or 3) to both for the variation of $d_6$ (**Figure 6a**) and $n_6$ (**Figure 6b**). Linear variation of dΨ and exponential variation of dΔ can be obtained at $\theta = 40.0°$, $\lambda = 522.5$ nm, where dΨ = 0.0630 - 1.1701$d_6$ and dΔ= 41.0173(1-exp(-1.2875$d_6$))$^{0.7544}$. There is very pronounced Ψ sensitivity at $\theta = 30.0°$, $\lambda = 533.5$ nm, with very small Δ variation: dΨ = -0.07186 + 2.39133$d_6$, and dΔ = 0.07829 + 0.10462$d_6$ + 0.78168$d_6^2$ - 0.80666$d_6^3$. High phase senility can be obtained at $\theta = 39.0°$, $\lambda = 524.0$ nm, where dΔ reaches 13° and 88° for 0.1 nm and 1 nm thin layer, respectively. It is almost 30 times higher dΔ variation in respect to the pure Si[6], where $n$ of the solution was about 1.34. In our case $n$ is about 1.5 (see PMMA, in the Figure 2). The higher is the difference of the refractive index of the detectable layer respect to the sensor effective $n$, the higher variation can be observed in (Ψ, Δ) spectra. Thus, for the multilayer sensor the dΨ and dΔ could be even larger if we would simulate PMMA with the smaller refractive index.

For biomedical sensor development, it is of high interest to detect the solutions of different concentrations, what can be seen as a variation of refractive index[6]. Both dΨ and dΔ has linear variation with refractive index $n_6$ (**Figure 6b**). The fitting parameters of the linear fit are summarized in the **Table 2**. The (40.0°; 522.5 nm) and (38.0°; 525.0 nm) is a relatively good compromise for double parameter sensing, while at (39.0°; 524.0 nm) a great sensitivity to phase shift, and at (38.0°; 525.0 nm) a higher sensitivity to the reflectivity measurements can be obtained. The slope of dΔ($n_6$) at (39.0°; 524.0 nm) is of 40$n_6$, what is slightly lower respect to pure Si[6] (see **Table 1**), but difference in detectable layer refractive index should be considered as just discussed above. Moreover, the aim of this paper was to develop sensors sensitive to both reflectivity and phase measurements, what is not the case of pure Si sensor[6].

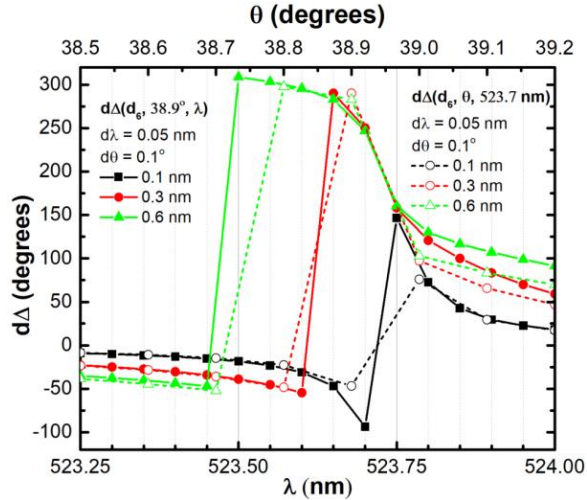

**Figure 5.** Phase shift difference dΔ as a function of thickness $d_6$ (polymer layer on the top of the multilayer sensor) and (38.9°, $\lambda$), and as a function of ($\theta$, 523.7 nm).

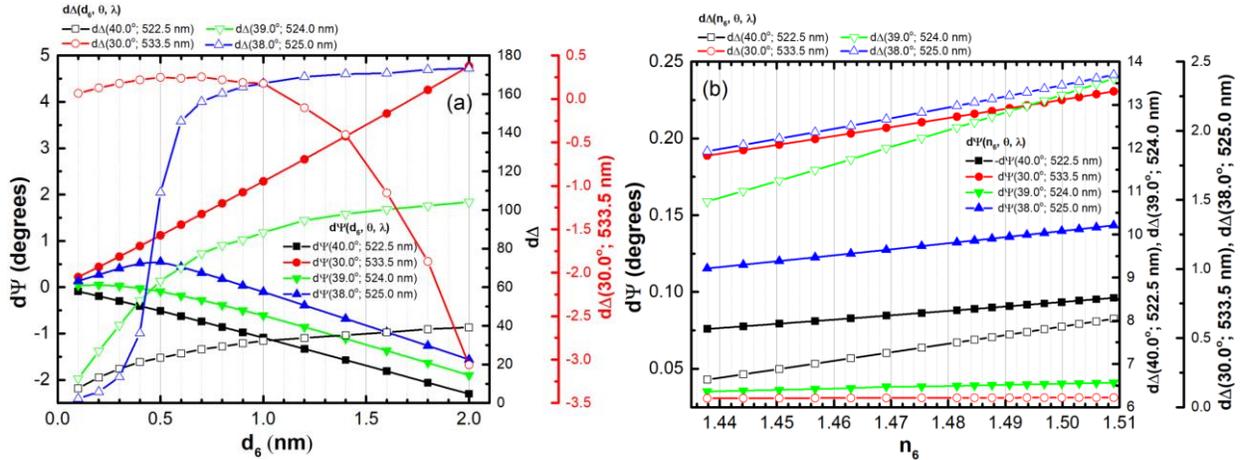

**Figure 6.** (a) The variation of (dΨ, dΔ) as a function of $d_6$ (a) and $n_6$ (b) for 0.1 nm thin $d_6$ thickness: four examples are given for different combinations of the ($\theta$, $\lambda$).

If the thickness $d_1$, $d_3$, $d_4$, or $d_5$ is increasing or decreasing down to zero, the sensor still has sensitivity in reflection measurements, but loses the sensitivity to the phase variation (**Table 3**). The increase in the thickness $d_2$ of YHO permits to shift the spectral region of the (Ψ, Δ) sensitivity.

**Table 2.** The fitting parameters of the linear fit dΨ = $A + Bn_6$ and dΔ = $A + Bn_6$ to the dΨ and dΔ functions given in the **Figure 6b**.

|  | dΨ | | dΔ | |
|---|---|---|---|---|
|  | A | B | A | B |
| (40.0°; 522.5 nm) | 0.33394 ± 0.0014 | -0.28494 ± 9.92383E-4 | -21.8751 ± 0.01641 | 19.8279 ± 0.01109 |
| (30.0°; 533.5 nm) | -0.65848 ± 0.002 | 0.58913 ± 0.00135 | -0.05249 ± 7.24999E-4 | 0.08098 ± 4.90017E-4 |
| (39.0°; 524.0 nm) | -0.08036 ± 5.7915E-4 | 0.08042 ± 3.9144E-4 | -46.79548 ± 0.19754 | 40.01705 ± 0.13351 |
| (38.0°; 525.0 nm) | -0.45124 ± 0.00155 | 0.39397 ± 0.00105 | -9.34342 ± 0.0694 | 7.78125 ± 0.0469 |

**Table 3.** The characteristics (behavior) of minimum values of $\Psi_{min}$ (zero or quasi zero reflection), $\Delta$ shift at corresponding $d\Psi_{min}$, difference in $d\Psi$ and $d\Delta$ with the variation (increase ↑, decrease ↓) of $d_1, d_2, d_3, d_4, d_5$.

|  |  | $\Psi_{min}$ and $d\Psi$ | $\Delta$ and $d\Delta$ |
|---|---|---|---|
| $d_1$ | ↑ | $\Psi_{min}$ is ↑ and slightly shift towards longer $\lambda$ | $\Delta$ ↑, $d\Delta$↓ and reaches 0° |
|  | ↓ | $\Psi_{min}$ is ↑ and slightly shift towards shorter $\lambda$ | $\Delta$ ↓, $d\Delta$↓ and reaches 0° |
| $d_2$ | ↑ | $\Psi_{min}$ ↑ up at ~5° and strongly shift towards longer $\lambda$; for the thickness from 10 - 50 nm the $\lambda$ of $\Psi_{min}$ moves between 515 - 660 nm; $d\Psi$ reaches ≈ -0.11° and ≈ 0.04° | $\Delta$ ↑ from ≈250° up to 350°; $d\Delta$ ≈ ± 40° |
|  | ↓ | $\Psi_{min}$ is ↑ and slightly shift towards shorter $\lambda$ | $\Delta$ ↓, $d\Delta$↓ and reaches 0° |
| $d_3$ | ↑ | $\Psi_{min}$ is ↑ and slightly shift towards longer $\lambda$ | $\Delta$ ↑, $d\Delta$↓ and reaches 0° |
|  | ↓ | $\Psi_{min}$ is ↑ and slightly shift towards shorter $\lambda$ | $\Delta$ ↑, $d\Delta$↓ and reaches 0° |
| $d_4$ | ↑ | $\Psi_{min}$ is ↑ and slightly shift towards longer $\lambda$, $d\Psi$ ≈ -0.12° | $\Delta$ ↑, $d\Delta$↓ and reaches 0° |
|  | ↓ | $\Psi_{min}$ is ↑ and slightly shift towards shorter $\lambda$, $d\Psi$ ≈ +0.12° | $\Delta$ ↑, $d\Delta$↓ and reaches 0° |
| $d_5$ | ↑ | $\Psi_{min}$ is ↑ and slightly shift towards shorter $\lambda$ | $\Delta$ ↓, $d\Delta$↓ and reaches 0° |
|  | ↓ | $\Psi_{min}$ is ↑ and slightly shift towards shorter $\lambda$, $d\Psi$ ≈ -0.25° | $\Delta$ maximum values remain the same, $d\Delta$↓ and reaches 0° |

The sensor can be simplified by removing the Layer 4, but the Layer 3 thickness must be increased up to 70 nm to keep the sensitivity to Δ: Si/Au (20 nm)/YHO (16 nm)/SiO₂ (70 nm)/Au (10 nm). For such sensor $d\Psi$ reaches -0.15° and $d\Delta$ reaches ±30.0° at $\lambda \approx$ 484.8 nm and $\theta$ = 40.5°. The $d\Psi$ reaches 0.23° at $\lambda \approx$ 492.5 nm and $\theta$ = 35.0°.

If low ($n$, $k$) YHO semiconductor is substituted with the zero-extinction coefficient (at visible range) semiconductor like $ZnO_x$ (**Figure 2**), the best conditions for the $d\Psi$ and $d\Delta$ sensitivity is at $\lambda \approx 515.8$ nm and $\theta = 32.5°$ with corresponding dΨ and dΔ 0.18° and ± 55.0°, respectively (the configuration: Si/Au (60 nm)/ZnO$_x$ (12 nm)/SiO$_2$(45 nm)/ZnO$_x$ (15 nm)/Au (20 nm)). If the Layer 4 is removed (configuration Si/Au (60 nm)/ZnO$_x$ (12 nm)/SiO$_2$(45 nm)/Au (9 nm)), dΨ reaches -0.30° and dΔ reaches ±10.0° at $\lambda \approx 471.6$ nm and $\theta = 32.5°$.

Various other simpler sensors can be designed (e.g., Si/Metal/LnkM/ Metal, Si/LnkM/Metal, LnkM/Metal/LnkM/Metal, where LnkM is any low ($n$, $k$) material like SiO$_2$, ZnO$_x$, PMMA, already reported elsewhere by other researchers[23,25,26]) to achieve similar or even greater reflectivity sensitivity. However, the phase shift reaching values up to 225° for only 0.1 nm thin detectable layer are unique and showed here for the multilayer of Si/Au (30 nm)/YHO (12 nm)/SiO$_2$ (35 nm)/YHO (15 nm)/Au (6 nm). The particularity of this configuration is based on low ($n$, $k$) YHO semiconductor material, ultra-thin thicknesses and gradual decrease of the refractive index towards SiO$_2$ layer and accordingly gradual increase of refractive index towards top Au layer for the electric field enhancement.

## 4. Conclusion

High reflection and phase shift sensitivity can be achieved by design of five-layer sensor (**Figure 1**), where $\widetilde{N_0} > \widetilde{N_1} > \widetilde{N_2} > \widetilde{N_3}$, $\widetilde{N_1} = \widetilde{N_5}$ and $\widetilde{N_2} = \widetilde{N_4}$. The highest variation of Δ up to 225° is observed at $\lambda = 523.7$ nm and at $\theta = 38.9°$ for the multilayer configuration of Si/Au (30 nm)/YHO (12 nm)/SiO$_2$ (35 nm)/YHO (15 nm)/Au (6 nm). This sensor has quasi zero reflection in the broad spectral range of (350 - 540) nm between $\theta = [30.0°; 70.0°]$. Linear variation of dΨ and exponential variation of dΔ can be obtained at $\lambda = 522.5$ nm and $\theta = 40.0°$ with variation of the detectable material thickness; at the same ($\lambda$, $\theta$) both (dΨ, dΔ) have linear variation with the detectable material refractive index change. The increase in the thickness $d_2$ of YHO permits to shift the spectral region of the (Ψ, Δ) sensitivity.

**ACKNOWLEDGMENTS**

This research is supported by the Horizon 2020 Project CAMART² under grant agreement No 739508 and national project "Thin films of rare-earth oxy-hydrides for photochromic applications" FLPP No LZP-2020/2-0291.